\documentclass[aps,prd,preprint,superscriptaddress,showpacs]{revtex4}
\usepackage{graphicx}

\usepackage{ulem}
\usepackage{color}
\definecolor{My_red}        {cmyk}{0.00,1.00,1.00,0.20}

\newcommand{\bmat}{\left(\begin{array}}
\newcommand{\emat}{\end{array}\right)}
\newcommand{\beq}{\begin{equation}}
\newcommand{\eeq}{\end{equation}}
\newcommand{\wt}{\widetilde}

\def\ra{\rightarrow}

\def\Ld{\Lambda}
\def\ld{\lambda}
\def\f{\frac}

\def\bwt{\begin{widetext}}
\def\ewt{\end{widetext}}
\def\be{\begin{equation}}
\def\ee{\end{equation}}
\def\bea{\begin{eqnarray}}
\def\eea{\end{eqnarray}}
\def\bean{\begin{eqnarray*}}
\def\eean{\end{eqnarray*}}
\def\bary{\begin{array}}
\def\eary{\end{array}}
\def\bit{\begin{itemize}}
\def\eit{\end{itemize}}

\def\ra{\rightarrow}

\def\Ld{\Lambda}

\def\ld{\lambda}

\def\su5u1{SU(5) \times U(1)}
\def\fsu5u1{SU(5) \times U(1)'}
\def\so10{SO(10)}
\def\sq20{SO(10) \times SO(10)}

\def\ra{\rightarrow}

\def\Ld{\Lambda}
\def\ld{\lambda}
\def\f{\frac}

\def\L{\left(}
\def\R{\right)}

\def\ra{\rightarrow}

\def\Ld{\Lambda}

\def\ld{\lambda}

\def\su5u1{SU(5) \times U(1)}
\def\fsu5u1{SU(5) \times U(1)'}
\def\so10{SO(10)}
\def\sq20{SO(10) \times SO(10)}

\usepackage[centertags]{amsmath}
\usepackage{amssymb}

\begin{document}

\title{Brightening the (130 GeV)  Gamma-Ray Line}

\author{Zhaofeng Kang}
\email{zhfkang@itp.ac.cn}
\affiliation{State Key Laboratory of Theoretical Physics and Kavli Institute for Theoretical Physics China (KITPC),
Institute of Theoretical Physics, Chinese Academy of Sciences, Beijing 100190, P. R. China}

\author{Tianjun Li}
\email{tli@itp.ac.cn}
\affiliation{State Key Laboratory of Theoretical Physics and Kavli Institute for Theoretical Physics China (KITPC),
Institute of Theoretical Physics, Chinese Academy of Sciences, Beijing 100190, P. R. China}

\affiliation{George P. and Cynthia W. Mitchell Institute for Fundamental Physics and Astronomy,
Texas A\&M University, College Station, TX 77843, USA}

\author{Jinmian Li}
\email{jmli@itp.ac.cn }
\affiliation{State Key Laboratory of Theoretical Physics and Kavli Institute for Theoretical Physics China (KITPC),
Institute of Theoretical Physics, Chinese Academy of Sciences, Beijing 100190, P. R. China}

\author{Yandong Liu}
\email{ydliu@itp.ac.cn}
\affiliation{State Key Laboratory of Theoretical Physics and Kavli Institute for Theoretical Physics China (KITPC),
Institute of Theoretical Physics, Chinese Academy of Sciences, Beijing 100190, P. R. China}

\date{\today}

\begin{abstract}

The gamma-ray line  from dark matter (DM) annihilation  is too weak  to  observe,  but its   observation will uncover     much information, e.g.,    the    DM mass  and an anomalously  large annihilation  rate $\sim0.1$ pb into di-photon. In this work, we construct a minimal effective theory (EFT) incorporating DM and heavier  charged particles. A large annihilation rate is obtained from  operator coefficients with resonance or strong coupling enhancement. The EFT is stringently constrained by the XENON100 and WMAP data.  Without resonance,  Dirac DM or  colored  charged particles are ruled out. It is pointed out that the  di-gluon mode   may correctly  determine the DM relic density.  Interestingly, this framework also provides an origin for  the Higgs di-photon excess at the LHC\@.  We apply the general analysis to the NMSSM, which  can elegantly  interpret the tentative  130 GeV gamma-ray line. A top-window model is also proposed to explain the gamma-ray line.
\end{abstract}


\maketitle

\section{Introduction and motivations}

The existence  of  dark matter (DM) has been confirmed by its gravitational effects, and its energy fraction $\sim25\%$ today is also measured. However,
the  conclusive   evidences that may reveal the DM particle properties are still absent. Among a variety    of  (indirect)   detecting objects on DM, the
gamma-ray from the DM  dense    region  (such as the center of the Galaxy) is especially promising by virtue of weak
astrophysical influence on its propagation~\cite{Bertone:2004pz}.  Of particular interest is the
monochromatic    gamma-ray line, which  has rather  clear background. But it is highly suppressed  because  the DM $\chi$ can only annihilate to photons  via the charged loop.

However, once  such a spectral  line is observed, it will uncover very important information  of DM\@.  In this article, we assume  an extracted   DM mass from  $E_\gamma$ and an anomalously  large annihilation rate $\langle \sigma v\rangle_{2\gamma}\sim0.1$ pb into di-photon   (it is taken as a referred value throughout the work, unless specified),
 then attempt to reconstruct the DM properties and dynmiacs  to the most extent. % with bias to natural particle physics  model.
  Inspired by the recent discovery of a gamma-ray line  at  $E_\gamma\simeq130$ GeV, which is claimed in the
  Ref.~\cite{Bringmann:2012vr,CW}
after  re-analyzing the    Fermi Large Area Telescope (FERMI-LAT) data published in 2009~\cite{Atwood:2009ez}, it is conjectured that the line may  originate from  DM annihilating into gamma. Best fit of the data shows  a DM of mass around $130$ GeV and  annihilation  rate at level  $0.1$ pb.
Later    independent analysis also confirms the line~\cite{Raidal}. The line has a  sharper peak which is hard to explain by FERMI-bubbles~\cite{bubble}, while  DM+DM$\ra \gamma\gamma$ gives a better fit~\cite{Raidal}. The Ref.~\cite{Su:2012ft} also shows  a strong evidence of the gamma-ray  from the inner galaxy and draws a similar  conclusion. This line has received much   attention from astrophysics~\cite{bubble,Raidal,Boyarsky:2012ca,astro} and   particle physics~\cite{particle}.
 In spite of
  queries~\cite{Boyarsky:2012ca},    the gamma-ray line from DM activity
 itself   is  of great  theoretical interest, and deserving a deep study.

The topic can be studied   in the  effective theory (EFT) framework, by minimally including an operator $a_C\chi^\dagger\chi C^\dagger C$ where $C$ is the charged particle.  The
anomalously bright   gamma-ray line is due to large $a_C$. We further demand the EFT be
compatible with  other constraints on the DM, i.e., the WMAP and XENON100 bound~\cite{XENON100}.
 Independent of the   mechanism  generating large $a_C$, we can arrive:
 \begin{itemize}
   \item The  charged  particle $C$ in the  loop should be  heavier than the DM, otherwise it would render
 too large annihilation rate into $C\bar C$, which leads to too small DM relic density. On top of that, the injection from
  such a large flux of charged particles into the cosmic-ray probably has been  excluded by the  PAMELA.
   \item  The charged particle carrying both QED and QCD charges needs careful inspections.
    Along with the di-photon annihilating  mode, there is an enhanced di-gluon mode  with estimated  rate
   $\langle \sigma v\rangle_{2G}\sim0.1(\alpha_s^2/\alpha^2)\langle \sigma v\rangle_{2\gamma}\simeq 1$ pb, which makes an illustrative  coincidence.
    %But quite  interesting, \emph{the di-$Z$ mode may accidently gives the right order, $\langle \sigma v\rangle_{2Z}\sim1$ pb}.
 \end{itemize}

The large  $a_C$ can be generated through   Breit-Weigner resonance mechanism, or the  strong  interaction between DM and the  charged loop. For the former scenario,  properties of the  scalar/vector resonance can be further stringently  restricted by  symmetries, e.g.\ the CP, and the above consideration. The next-to-minimal supersymmetric standard model (NMSSM)~\cite{Ellwanger:2009dp} is a good realization of this scenario. We find it is capable of interpreting  the
tentative 130 GeV gamma-ray line. For the latter scenario, the XENON100 bound excludes the Dirac DM, as well as both Dirac and Majorana DM if the charged paticle $C$ carries
color. Interestingly, in any scenario, the possible  SM-like  Higgs $h$ to di-photon excess at the LHC~\cite{LHC:Higgs}  may share the same origin, if we incorporate the
operator $a^h_C hC^\dagger C$.

This paper is organized as following:  In the section~\ref{2},  we perform  a general analysis based on the minimal EFT. In the next two sections exploration on   the enhancement mechanism is presented. The Section~\ref{conclusion} includes  the conclusion and  discussions.
 And some necessary complementarity is casted in the Appendix. %Final section gives the conclusion and discussion.

\section{Generality  and guidance}\label{2}

As is well known, the DM can not directly annihilate into photons due to its QED charge neutrality, while  transition at the loop level is generically  highly suppressed.
 So, it is nontrivial to obtain  an abnormally large annihilating rate, saying  $\langle\sigma v\rangle_{2\gamma}\sim0.1$ pb.
 Some more powerful model independent  statements can be made, if it is further combined with other aspects of DM. To see that, we consider the minimal  effective operators~\footnote{Obviously, our effective discussion is still  more fundamental than the Ref.~\cite{Rajaraman:2012db} that only give operators consists of
 DM field  and gauge/Higgs fields. And it is can be a generation  of~\cite{Goodman:2010qn} that only includes light charged particles.} relevant to the gamma-ray line anomaly% (closing the charged loop determines the $\langle\sigma v\rangle_{2\gamma}$)
\begin{align}\label{effective:1}
{\rm Fermion \,\,DM}:&\quad \quad a_C\bar\chi\Gamma\chi\bar C\Gamma C,\quad a_{\wt C}\bar\chi\Gamma\chi {\wt C}^\dagger {\wt C},
\quad a_{W}\bar\chi\Gamma\chi W^+W^-,\\
{\rm Scalar \,\,DM}:&\quad \quad a_t\chi^\dagger\chi\bar C \Gamma C,\quad\quad\,\,\, a_W\chi^\dagger \chi W^+W^-,
\end{align}
where the gamma matrix  $\Gamma\subset\{1,\gamma^5,\gamma^\mu,\sigma^{\mu\nu}\}$. Lorentz and $SU(3)_C\times U(1)_{\rm QED}\times$ CP symmetries are implied,  and some operators will vanish due to these symmetries.
 $C$ is  a charged fermion and $\wt C$ is a charged scalar, both of which are not confined to  the SM. However,  in the sense of inducing DM annihilating into gamma,
 the scalar loop is not as effective as the fermionic loop, unless there is a large enhancement from highly charged particles such as a double charged scalar.
  Note that operators containing  $H^+W^-$ where $H^+$ is the charged Higgs from 2HDM-like model are not included, since their contribution are always putative null, $e.g.$, in the   non-linear unitary gauge the vertex $H^+W^-\gamma$ vanishes~\cite{Bergstrom:1997fj}.

 To achieve a large $\langle\sigma v\rangle_{2\gamma}$ and maintain the main merit of DM dynamics, the candidates running in the charged loop are more or less selected.
Denoting  by  $S_{i}$ the set of
 DM $2\ra2$ annihilation  mode
 and without loss of generality, let  $S_1$  be the one from which di-photons  come after closing the charged states to form a loop. Some cases arise:
 \begin{itemize}
   \item If $S_1$ is the on-shell type with final states $X_1\bar X_1$, then $\langle\sigma v\rangle_{S_1}\sim10^{4}\langle\sigma v\rangle_{S_1,2\gamma}\sim10^3$ pb. Injection from such a large flux charged particles would have been observed by PAMELA~\cite{PAMELA}  from the significant  excess of the positron  or anti-proton flux. On top of that, the DM relic density would be too small, unless  we consider the subtle thermal Breit-Weigner    enhancement effect which is active only today~\cite{Breit}.
   \item If $S_1$ is properly off-shell,  then the above problem is resolved since the $S_1$ is forbidden today.
   But  we have to examine its annihilation rate at the  early universe, $i.e.$,  comparing   the rate of the forbidden annihilation  mode
   $\langle\sigma v\rangle_{S_1,T_f}$ with 1 pb, where $T_f= m_\chi/x_f$ with $x_f\simeq 25$  the typical  decoupling temperature of DM.
 Generically one can expand  the annihilation rate as~\cite{threeEX}
    \begin{align}\label{}
\L \sigma v\R_{S_1}=(a+b/x_f)v_2,
\end{align}
where the final  two-body phase space  gives the velocity of  out-going particles in the CM frame,  $v_2=\L 1-z^2+z^2v_{rel}^2/4\R^{1/2}$ with  $z=m_{X_1}/m_\chi$ and $v_{rel}$ the relative velocity of initial particles. For a properly large $z>1$, the relic density of DM is~\cite{threeEX}
   \begin{align}\label{forbidden}
\Omega h^2=\f{1.07\times10^{9}x_f}{g_*^{1/2}M_{\rm Pl }J},\quad J\simeq a\f{z}{\mu_-x_f}e^{-\mu_-^2x_f},
\end{align}
  where $\mu_-=(1-1/z^2)^{1/2}$.  The relic density is very sensitive to $z$, $e.g.$, from $z=1.05$ to $z=1.10$, it increases  roughly one order.
  But in principle it is possible to obtain $\langle \sigma v\rangle_{2\gamma}\sim0.1$ pb and $\langle \sigma v\rangle_{X_1\bar X_1}\simeq 1$ pb simultaneously, if we accept large fine-tuning.
   \item  If     $z$ is large enough  then the forbidden channel is completely ignorable.
    As a consequence,   we need a new (dominant) channel $S_2$ to reduce the DM number density. In actual model building, it  naturally happens. But a more interesting  case arises as following.
\item If the charged particle also  carry  color charge, then the $2\gamma$ mode is subdominant to the two gluon mode:
  \begin{align}\label{}
\f{\langle \sigma v\rangle_{2G}}{\langle \sigma v\rangle_{2\gamma}}\sim0.1\f{\alpha_s^2}{\alpha^2}\simeq{\cal O}(20),
\end{align}
which is estimated in light of fermions with unit charge, and the origin of  numerical factor 0.1 can be traced back to  the property of charged particles.  This numerical coincidence
 means that if the two-gamma rate is $\sim 0.1$ pb, the right relic density is achieved via the di-gluon mode.
 For the proof of that the di-$Z$ mode is at most the same order of di-photon mode and thus irrelevant, see the Appendix~\ref{phigamma2} for details.
 \end{itemize}

Generically, we will have several annihilation  modes producing  gamma line,  $\chi\bar\chi\ra \gamma\gamma,\,\gamma X$ with $X=Z,h$. The first mode creates a line at $E_\gamma=m_\chi$ while the second mode produces a line  with lower energy  $E_\gamma'\approx E_\gamma(1-m_{X}^2/m^2_{\chi})^{1/2}$.
 They have comparable cross sections except for a very significant phase space suppress.
Therefore we focus on the two gamma final states, since the two modes quantitatively differ only by some constant, such as the difference between $e$ and $g_2$.
 However,  how to distinguish  the  two lines is very interesting as discussed in
Ref.~\cite{Rajaraman:2012db}.

To end up the general discussion in  EFT, we would like to mention that there is a possible relation between the di-phonon excess for the Higgs search at the CMS/ATLAS~\cite{LHC:Higgs} and for the DM search at the sky.
The   common point is the new charged loop. Through the same loop alone  which the  DM annihilates into two photons, the SM-like  Higgs $h$ can decay into two photons with appreciable width, if the coupling to the Higgs is
 significantly. It is can be described simply by further including the  effective operators   $a_C^hh\bar CC$ or $a_C^hh \wt C^\dagger \wt C$.
However we are not going to discuss this in detail due to its triviality in the EFT.

\section{Enhancement from  Resonance}

In this section, we present the effective analysis by specifying the role of resonance. The  simple top window  model is constructed,
 and in particular we survey   its  implication on   conventional supersymmetric model such as the  NMSSM.

\subsection{Scalar resonance}\label{res:scalar}

We consider the scalar resonance which appears almost everywhere in models with extended Higgs sector. To get a  sufficiently  large enhancement, $s-$channel resonant annihilation is the most conventional mechanism. In this  case,  the
resonance $\phi$ takes mass  $m_{\phi}\simeq2m_{\rm DM}$, and  the cross section manifest of the enhancement  can be parameterized as
  \begin{align}\label{}
\sigma v=\f{T_IT_F}{32\pi}\f{1}{m_\chi^2}|{\cal M}|^2\sim\f{\alpha^2 m_\chi^2}{m_{\phi}^4}\f{1}{\L1-r\R^2+\gamma},
\end{align}
where $r=4m_{\chi}^2/m_\phi^2$ and $\gamma=(\Gamma_\phi/m_\phi)^2\ll1$.  Here  $\alpha$ stands for an effective coupling and will be specified in concrete examples,  while  $T_{I,F}$ takes $1/2$ or $1/4$ and so on, standing for   the average of initial  degree of freedoms  or the  symmetry factor of final states.

It is  convenient to   define $f_B=1/(\L1-r\R^2+\gamma)$, then  the DM annihilating  cross section into  $X_i\bar X_i$ (thought $\phi$) can be rewritten as
%We can make a model independent constrain on the Br$(\phi\ra X_2\bar X_2)$:
        \begin{align}\label{phi:11}
(\sigma v)_{S_i}=\f{f_B}{m_\phi^4}\f{T_I}{m_\chi}|{\cal M}(\chi\chi\ra\phi)|^2\,\Gamma(\phi\ra X_i\bar X_i).
\end{align}
Without loss of generality, we take $X_2$ as the dominant mode, and immediately get the upper bound of the branching ratio of  $\phi$ decay (to particles other than di-photon):
        \begin{align}\label{phi:1}
\f{{\rm Br}(\phi\ra X_2\bar X_2)}{{\rm Br}(\phi\ra \gamma\gamma)}=\f{(\sigma v)_{S_2}}{(\sigma v)_{2\gamma}}\lesssim 10.
\end{align}
 To  arrive it we have set  $(\sigma v)_{S_2}\sim1$ pb as the standard annihilation rate as well as  the referred value $(\sigma v)_{2\gamma}\sim0.1$ pb.  Therefore we get   a model independent bound ${{\rm Br}(\phi\ra \gamma\gamma)}\gtrsim10\%$, as negates the resonance from simple two-Higgs-doublet-model (2HDM) by virtue of their considerably  coupling to fermions  or light massive vector boson (it is absence for CP-odd Higgs).
This   bound  has far-reaching implication on the collider. Provided that the production cross section of $\phi$
is sufficiently large (for example, when  the charged loop meanwhile carries color  as in the top-window model discussed later),  \emph{the gamma-ray line observed at the sky  predicts a clear di-photon excess at the peak around 2$m_\chi(\approx260$ GeV) at the LHC}.

\subsubsection{Effective analysis}\label{topW}

In  light of previous arguments, we need some rather heavy charged particles, while
in the SM  top quark and $W$ boson are the only two charged particles of mass around the weak scale. Accordingly, a $m_\chi>m_t$ hints a new charged particle. When  $m_t>m_\chi>m_W$, the top quark  will  open a unique window. Otherwise, $W$ may run in the charged loop provided a vertex $\phi W^+W^-$, that implies nontrivially the identity  of $\phi$.

Since the  effective operators listed in the previous section are ascribed to the integrating out $s-$channel scalar  resonance,
 the set of  possible operators can be reduced greatly:
\begin{align}\label{effective:1}
{\rm Fermionic \,\,DM}:&\quad \quad a_C\bar\chi(\gamma^5)\chi\bar C(\gamma^5) C,\quad a_{\wt C}\bar\chi(\gamma^5)\chi {\wt C}^\dagger {\wt C},
\quad a_{W}\bar\chi\chi W^+W^-,\\
{\rm Scalar \,\,DM}:&\quad \quad a_C\chi^\dagger\chi\bar C (\gamma^5) C,\quad \quad a_{\wt C}\chi^\dagger \chi {\wt C}^\dagger {\wt C},\quad\quad\,\,\, a_W\chi^\dagger \chi W^+W^-.
\end{align}
%Keep in mind that
The
%resonantly enhanced operator
coefficients $a$'s are proportional to $f_B^{1/2}/m_\phi^2$. And
 $\gamma^5$ in the parenthesis  may or may not appear, depending on the CP quantum number of $\phi$.
%On the other hand, specified to
But specified to    fermionic DM,   $\gamma^5$ must be inserted so as to make the present DM  annihilating  rate   avoid acute velocity suppressing, $v^2\sim10^{-6}$.
This means the $\phi$ must be CP-odd, denoted as $\phi_A$ hereafter. As an immediate  consequence,   the $W-$window is closed. Additionally,  $\wt C^\dagger\wt C$  should be understood as $\wt C_L^\dagger\wt C_R$, under CP transformation
$\wt C_{L/R}^\dagger\ra \wt C_{R/L}$. However,   QED does not change the chirality,  thus we  need a further large LR mixing.  As an example, the stop-system   in the SUSY just satisfies   those requirements.
On the contrary, for scalar DM, no matter complex or real, the $\phi$ should be CP-even and denoted as $\phi_h$. Hence the $\gamma^5$ should be removed.

We would like to add some further remarks. Firstly, the Lorentz and SM-gauge invariance force  $\phi$ either transforms   non-trivially under the $SU(2)_L$ symmetry or mixes with such states. Secondly, the constraint on the $\phi$ interactions  indicated by the Eq.~(\ref{phi:1}) should be satisfied.
 Finally,  the $\phi$ also mediates the  tree-level
 DM-nucleon interaction in the presence  of a top window. Although this contribution is    suppressed by velocity for the fermionic DM, the scalar DM requires inspection~\cite{Gao:2011ka}. The resulting DM-proton inelastic scattering cross section   is
    \begin{align}\label{top:DD}
\sigma_{\rm SI}=\f{4\mu_p^2}{\pi}f_p^2,\quad f_p\simeq\f{2}{27}f^p_{T_G}\f{a_t}{2m_\chi}\f{m_p}{m_t},
\end{align}%\f{g_{\phi CC}g_{\phi\chi\chi}}
where $\mu_p\approx m_p$ is the DM-proton reduced mass and $f^p_{T_G}\simeq0.83$~\cite{Gao:2011ka}. Note  the enhancement factor $f_B^{1/2}$ in $a_t$ is removed when we are   calculating  the DM-nucleon recoil rate using the effective  operator.
 The present  exclusion on $f_p$ is $10^{-8}\rm\,GeV^{-2}$ for DM of mass 100 GeV, put by the XENON100~\cite{XENON100}. It implies the upper bound
    \begin{align}\label{scalar}
a_t\lesssim 3.2 f_p m_\chi m_t/m_p\simeq 2.8\times10^{-3}\L\f{f_p}{10^{-8}\rm\,GeV^{-2}}\R\L\f{m_\chi}{100\rm\, GeV}\R{\rm\,GeV^{-1}},
\end{align}%\f{g_{\phi CC}g_{\phi\chi\chi}}
which  places a rather strong constraint for the top-window model.

Having  outlined     the most essential profile  of  WIMP that potentially has bright gamma-ray lines, we continue  to make some quantitative discussion.
  Effectively, through  the  charged loop,    Lorentz and CP invariance leads to the  following operators for the    CP-even and CP-odd  $\phi$ respectively
  \begin{align}\label{phi2gamma}
\alpha\f{ h_{\phi CC}}{4\pi}\f{1}{4\Ld_1} \phi_h F_{\mu\nu}F^{\mu\nu},\quad
\alpha\f{ h_{\phi CC}}{4\pi}\f{1}{8\Ld_2} \phi_A F_{\mu\nu}\wt F^{\mu\nu},
\end{align}
with  $\alpha\approx1/137$.   $h_{\phi CC}$ is the coupling constant  between $\phi$
 and charged particles.
 Factoring  out the loop factor and couplings,  moreover multiplying   $1/4(8)$ for later convenience, the $\Ld_{1,2}$ can be much below the weak scale (it is even  enhanced by color or electric  charge). Concrete expressions for the effective scale are casted in the Appendix~\ref{phigamma2}.

 Now we are at the position to evaluate  the DM annihilation rate into  gamma pair.  Denoting  by $\Gamma_{\phi\gamma\gamma}^{\mu\nu}$ the Feynman  rules  (see Fig.~\ref{Fig:abc} for label)
 of  Eq.~(\ref{phi2gamma}),
 allowing for  off-shell  $\phi$,  % expressed in the momentum space , the
 they are   respectively given by~\footnote{In the  Feynman  rule of $\phi FF$  vertex, we have make it explicitly conserves the Ward-Takahashi identity by symmetric the $p_1^\mu p_2^\nu$ term.}
  \begin{align}\label{}
{\rm\phi\,\,\,CP-even:}&\quad\Gamma_{\phi\gamma\gamma}^{\mu\nu}(p_1,p_2,P)=\f{\alpha h_{\phi CC}}{4\pi}\f{1}{\Ld_1}\left[p_1\cdot p_2g^{\mu\nu}-\L p_1^\mu p_2^\nu+p_2^\mu p_1^\nu\R\right],\\
{\rm\phi\,\,\,CP-odd:}&\quad \Gamma_{\phi\gamma\gamma}^{\mu\nu}(q_1,q_2,P)=\f{\alpha h_{\phi CC}}{4\pi}\f{1}{\Ld_2}\epsilon^{\mu\nu\alpha\beta}(p_1)_\alpha  (p_2)_\beta.
\end{align}
After averaging  initial states  and summing over out-going    photon polarization states, the resulting  real scalar DM annihilation rate is
   \begin{align}\label{}
(\sigma v)_{2\gamma}\simeq\f{f_B}{128\pi }\f{g_{h\chi\chi}^2}{\Ld^2_1}\f{\alpha^2h_{\phi CC}^2}{16\pi^2}\f{1}{m_\chi^2}\approx0.06\L\f{f_B}{500}\R\L\f{30\rm\,GeV}{\Ld_1}\R^2\L\f{g_{\phi\chi\chi}'h_{\phi CC}}{0.3}\R^2\rm\,pb,
\end{align}
%����ħ��=m_\phi^4/2\Ld^2 \f{f_B}{m_\phi^4} ĩ̬����ȫͬ����̬��ƽ�������Գ���2*2=4.
%We have took DM as self-conjugate particle, or else the estimated  value should be half.
where  $g_{h\chi\chi}$ is the coupling constant  between $\phi$ and DM.  Owing to the scalar DM,  we have   parameterized  the massive coupling  as $g_{\phi\chi\chi}\ra 2g_{\phi\chi\chi}'m_\chi$. For the Majorana  DM,
   \begin{align}\label{2gamma:scalar}
(\sigma v)_{2\gamma}\simeq\f{f_B}{128\pi }\f{g_{h\chi\chi}^2}{\Ld^2_2}\f{\alpha^2h_{\phi CC}^2}{16\pi^2}\approx0.04\L\f{f_B}{100}\R\L\f{30\rm\,GeV}{\Ld_2}\R^2\L\f{g_{\phi\chi\chi}h_{\phi CC}}{1.0}\R^2\rm\,pb.
\end{align}

%%%%fig.2%%%%%%%%%%%%%%%%
 \begin{figure}[htb]
\begin{center}
\includegraphics[width=3.0in]{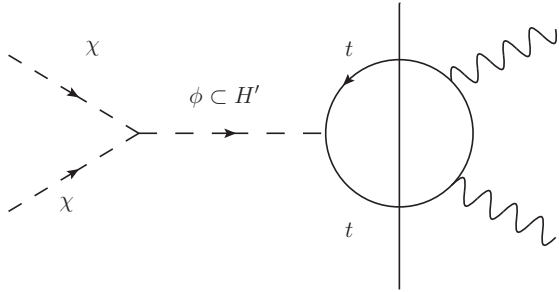}
\end{center}
\caption{\label{2gamma} Dark matter annihilates into  $2\gamma$ via top loop. The cut denotes for annihilates into top pair.}
\end{figure}
%%%%%%%%%%%%%%%%%%%%%%%%%%%%%%%%%%%%%%%%%%%%%%%%%%%%%%%%%%%%%%%%%%%

 %the Higgs di-photon excess maybe interpreted.

\subsubsection{The top-window model}\label{}

In the standard model (SM), lying on the top of the fermion  mass ascending order,  the   top quark may hide some new dynamics. It is thus interesting  to conjecture that the  dark sector may have a close relation with the top quark, $i.e.$,  the dark sector  only  (strongly) interacting  with top quark in  the fermion sector~\footnote{We note that alone this line, a pioneer attempt has been made in~\cite{Jackson:2009kg}, in which  a vector resonance is investigated.}.
 Then it is reasonable to expect enhanced gamma-ray line via top-loop  at levels close to the present experimental sensitivity.
And interesting, the latest  tentative gamma-ray line from dark matter annihilation~\cite{CW} requires a DM of mass about 130 GeV, just lies within the top-window.
Therefore, as a concrete example  of the general  effective operator analysis,  we focus on a top-window model equipped with scalar resonance.

Due to the chiral  structure of the SM,  it is most likely that  the $\phi$ dwells in an extra Higgs doublet $H'$ with hypercharge $+1/2$ so as to couple the top quarks at dimension-four level. A simple  effective model, which  might be  a remnant of top dynamics reads
\begin{align}\label{}
{\cal L}&=\f{1}{2}\chi^2\L\mu_\chi^2+\ld H'H+c.c.+\kappa_1|H'|^2+\kappa_2|H|^2\R+V(H,H')+y_{3}\bar Q_{3} H' t_R.
\end{align}
where  $H$ is the ordinary   Higgs doublet developing a vacuum  expected  value (VEV) $v=174$ GeV. By contrast, $H'$ should have a  VEV $\langle H'\rangle\ll v$ which makes the term   $\phi W^+W^-$   neglectable~\footnote{Such a spectator Higgs doublet is  motivated   in the Ref.~\cite{Gao:2011ka} to explain the origin of   iso-spin violation between  the  DM-nucleon interaction.}. Here  we  assume that the  VEV hierarchy is realized  by well organized Higgs potential $V(H,H')$ as discussed in~\cite{Gao:2011ka}. In addition, since we have ascribed the unique significant   coupling  $H'\bar Q_{3}t_R$    to some unknown dynamics, we are free of the FCNC-problem induced by family changing  Yukawa couplings $H'\bar f_Lf_R'$.
 In summary, after the EW-breaking, we are left with  the relevant terms
\begin{align}\label{}
{\cal L}\supset  \f{1}{2}\L\mu_\chi^2+\kappa_2v^2\R\chi^2+ \f{1}{2\sqrt{2}}(\ld v)\chi^2\phi+ \f{1}{\sqrt{2}}(\kappa_2 v)\chi^2h+\f{\kappa_2}{4}\chi^2h^2+\L \f{y_3}{\sqrt{2}}\phi\bar t_Lt_R+c.c.\R.
\end{align}
In  light of the general analysis, the  scalar DM the
  $\phi$ is identified with the CP-even component of the neutral boson in  $H'$. If we  drop    terms involving  $H'$, the model just  recovers the Higgs-porting model (see a recent discussion~\cite{Djouadi:2012zc}). But since $\kappa_2$ is irrelevant on our purpose, so we can  turn it off to reduce parameters.

  %   Notice that the $\ld$ term is crucial to produce
  %the    $\chi^2 \phi$ vertex, so it is can not be forbidden by $Z_2$-symmetry acting on  $H'$ and dark matter.
  % The presence of such a  term   excludes the assignment of  on so as to naturally suppress  $\langle H'\rangle$.

Now we turn our attention to the phenomenological aspects of the model. First of all, the di-photon rate, in terms of the parameterization in the Eq.~(\ref{2gamma:scalar}), is determined by
    \begin{align}\label{}
g_{\phi\chi\chi}'=\f{\ld}{2^{3/2}}\f{ v}{m_\chi},\quad h_{\phi CC}=y_3/\sqrt{2}.
\end{align}
Taking $\ld\simeq y_3\simeq1$ and $f_B=500$ leads to a rate $\simeq$0.07 pb. And the XENON100 constraint Eq.~(\ref{scalar}) has been arranged to be  satisfied by means of rather large  $f_B$ thus smaller couplings.
Next,   the mass difference between
$m_t$ and $m_\chi$ is at a few ten percents level, so the forbidden annihilation mode (to top quark) is ignorable in terms of the Eq.~(\ref{forbidden}).
The di-gluon mode may properly  account for the relic density. From  Eq.~(\ref{Lds}) and  Eq.~(\ref{QEDLd}) it is estimated that
  \begin{align}\label{}
\f{\langle \sigma v\rangle_{2G}}{\langle \sigma v\rangle_{2\gamma}}=\f{\alpha_s^2}{\alpha^2}\f{1}{4N_c^2Q_t^4}\approx35,
\end{align}
 with $Q_t=2/3$ the top quark charge.

An alternative model is to assume the dark sector consists of a SM vector-like  Dirac fermion pair $(\psi,\psi^c)$, which is the
 the $SU(2)_L$ doublet and
carries
hypercharge $\pm1/2$ respectively.  Extra singlet fermion $S$ is introduced, then the model is
\begin{align}\label{}
{\cal L}=\L \ld S\bar\psi H'+m_\psi\psi\psi^c+M_SS^2\R+V(H,H')+y_{3}\bar Q_{3} H' t_R.
\end{align}
The fermonic DM brings  important difference. The $\phi$ should be the CP-odd component of $H'$, which does not couple to $W^+W^-$  thus allows a large  $\langle H'\rangle$ (but we have to ensure small $H'-H$ mixing, otherwise  $\phi\bar bb$ is  too large). It leads to a phenomenologically viable  singlet-doublet mixing Majorana DM.
%We do not go to the detail of this model, since
 It can be regarded as an effective model of the supersymetric model studied as the following (but top loop will be  replaced by chargino loop).

%Such model is somewhat natural since the $H'$ can be assigned $Z_2$ symmetry, while coupling with top quark is the soft breaking term at the visible sector.  It explains the small VEV of $H'$.  Moreover, its detectable. We will postpone the detail of this model, since it has a natural realization in the
% SUSY with replacement top as charginoo.

\subsection{The NMSSM}

In the supersymmetric standard  models (SSM), in addition to the (Majorana) neutralino  LSP dark matter of mass  around $m_Z$,
a wealth of  new heavy charged particles,    CP-odd Higgs and hence a viable resonance $\phi_A$,   are furnished. It is thus of particular  interest to investigate
 whether  the above set up can be realized in the SUSY. Furthermore,  as stressed at the beginning of this section, the coupling of the resonance to light states are stringently constrained. In spite of  difficulties in the minimal-SSM, \emph{such an invisible $\phi_A$ can be readily  accommodated in the  NMSSM following  the  singlet limit}.
 While   the examination on light dark matter limit has been done~\cite{Chalons:2011ia}, our scenario has  not been currently examined yet as far as we are aware.

%%%%fig.2%%%%%%%%%%%%%%%%
 \begin{figure}[htb]
\begin{center}
\includegraphics[width=3.8in]{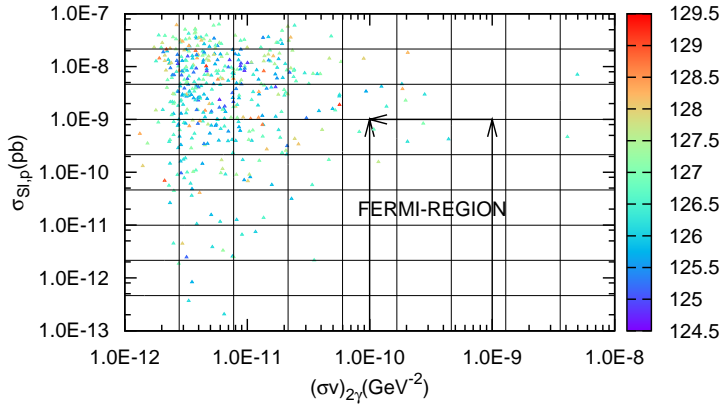}
\end{center}
\caption{\label{FERMI} The  LSP (with acceptable  relic density 0.09-0.12) at the plane of $(\sigma v)_{2\gamma}-\sigma_{\rm SI}^p$. As one can see, the FERMI gamma-ray line can be accommodated even under stringent XENON100 exclusion. The colored notation
 is of the SM-like Higgs  mass. We obtain the results by using the  programme NMSSMtools~\cite{NMSSMTools}.}
\end{figure}
%%%%%%%%%%%%%%%%%%%%%%%%%%%%%%%%%%%%%%%%%%%%%%%%%%%%%%%%%%%%%%%%%%%

Owing to the  singlet sector, the NMSSM  presents  a clear realization of our scenario. For our purpose,  the  model  is none other than the $Z_3-$NMSSM:
\begin{align}\label{}
W&\supset\ld SH_uH_d+\f{\kappa}{3}S^3,\\
-{\cal L}_{soft}&\supset m_S^2|S|^2+\L \ld A_\ld SH_uH_d+ A_\kappa\f{\kappa}{3}S^3+c.c.\R,
\end{align}
Consider a slice of the parameter space: (A) $\ld\sim1$ moreover $\kappa\lesssim\ld$ and $\tan\beta\sim 3$ (favored by naturalness to enhance the SM-like Higgs mass~\cite{Kang:2012tn}), further  the $v_s\equiv\langle S\rangle$ gives rise to the  Higgsino-like charginos (with mass roughly  $\mu=\ld v_s\sim200$ GeV)
  which replaces the top quark in the gamma loop;
(B) The 130 GeV LSP dark matter  has significant  (even if not dominant)  singlet component,
and  $\kappa S^3/3$ provides the vertex $\kappa A_s \wt S^2$ with unsuppressed  coupling.
 %in practice the LSP in the parameter configuration under consideration is well mixed and one may have to properly tune the fraction to avoid XENON100 exclusion;
  (C) The highly  singlet-like CP-odd Higgs $A_1\simeq A_s$ of mass around  260 GeV offers a proper resonant enhancement.
Consequently   $A\bar bb$ and $AhZ$ can be sufficiently suppressed, and  its only significant coupling is to the chargino $\sim\ld A_1\wt H_u^+\wt H_d^-$.
This hiding $A_1$ is a key to reconcile the WMAP and FERMI, and the situation of this parameter space to interpret FERMI is shown in the Fig.~\ref{FERMI}.
We have restrict $m_{A_1}$ falls into 255-265 GeV. As one can see, only a very small portion of the points pass all constraints, labeled as the FERMI-region.

\begin{figure}[htb]
\begin{center}
\includegraphics[width=3.2in]{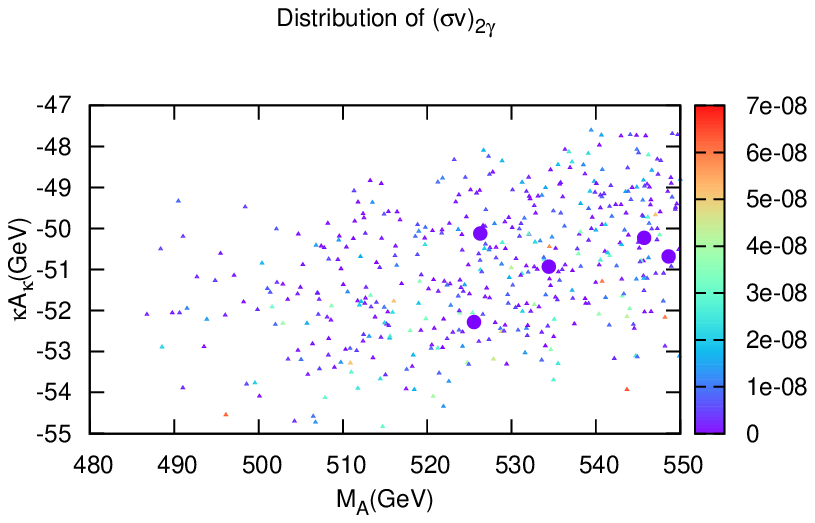}
\includegraphics[width=3.2in]{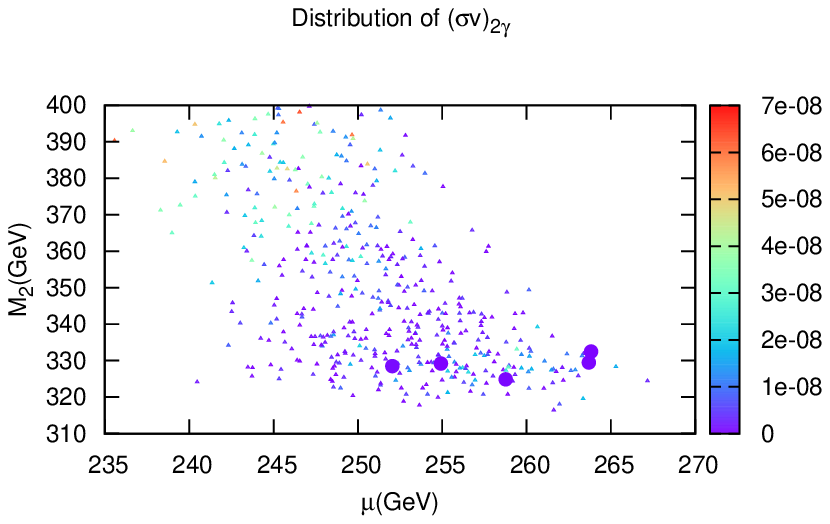}
\end{center}
\caption{\label{temp} The distribution of the di-photon  rate on input coordinates, left: $M_A-\kappa A_\kappa$; Right: $\mu-M_2$. The
solid purple  circles stand for parameter configurations satisfying  WMAP, XENON100 and FRRMI. They are well-tuned, since they   are rather discrete even in the preferred window.
 Other parameter settings: $\ld=0.65,\,\kappa:0.13-0.16,\,\tan\beta:1.3-1.7,\,m_{\wt Q_3}=m_{\wt t_R}=1000$ GeV, $A_t=0$ while $A_\kappa$ varies between  -400 to -300 GeV.}
\end{figure}

We  close this section by making some further comments.  The above parameter space is a portion of the natural NMSSM~\cite{Kang:2012tn}. However, as observed there, the Higgsino generically occupies  a large proportion of  LSP, and its relic density is too small while $\sigma_{\rm SI}^p$ is too large.
 To circumvent those problems,
  we may have to tune the parameters to obtain a viable LSP, as is reflected in the Fig.~\ref{temp}.
     From   model building, we  may simply go to the singlet-port dark sector by simply  adding  $\eta S\Phi^2/2$, where
the lighter   $\Phi$ component  is the dark matter candidate.

%%%%fig.2%%%%%%%%%%%%%%%%
%%%%%%%%%%%%%%%%%%%%%%%%%%%%%%%%%%%%%%%%%%%%%%%%%%%%%%%%%%%%%%%%%%%

\subsection{Vector resonance}

In this subsection we turn our attention to the vector resonance. There is only one neutral massive vector boson in the SM as well as its simple extension.
 Such a resonance means the   DM mass should be around $m_Z/2$. However, in   models with  extended   $U(1)_X$  local symmetries,  new resonances with free masses   are expected. Such models have been proposed in~\cite{Jackson:2009kg,2gammaV}, where
  spectral lines from $\gamma Z$ and/or $\gamma h$ final states  are predicted.
It is believed that the Landau-Yang theorem~\cite{CNY} excludes the di-photon mode. Nevertheless,
%states,  a spin-1 particle decay to two gamma is forbidden. Nevertheless,
this theorem does not exclude the vertex $Z\gamma\gamma$ with off-shell $m_Z$, please see the Ref.~\cite{WWV} discussing anomalous  three vector boson coupling within  SM. And a concrete evidence for
such coupling can be found in the earlier calculation of neutralino annihilation to di-photon~\cite{Bergstrom:1997fj}, where  a pole from $m_Z$ indeed exists.

This pole has deep relation with the anomaly of the theory, and it is non-vanishing if and only if the  axial-coupling between $Z'$ and the charged massive  fermions is present (mass splitting is needed to spoil thorough   anomaly cancellation).
%ʸ���������ʵ��Furry����֤��Ϊ�㡣
As a consequence, in  Eq.~(\ref{effective:1}) we are left with the only one  operator built from fermionic DM and charged particles:
\begin{align}\label{}
a_C\bar\chi\gamma^5\gamma_\mu\chi\bar C\gamma^5\gamma^\mu C,
\end{align}
which indicates  $Z'$ can not come from vector-like theories. While for the complex scalar DM,  the relevant effective vertex is $\chi^\dagger\tensor{\partial}^\mu\chi Z_\mu'$, which renders the DM annihilation suffering from velocity suppress. So we do not need to consider it here.

 %Similarly we start from the effective operators, but this time only the fermionic DM is under consideration. The reason is that And then from
%Specified to the Majorana DM that only participates the axial-vector coupling, we must include a $\gamma^5$ for bilinear DM.

Whatever the charged particles are, the $Z'\gamma\gamma$ effective Lagrangian can be built by Lorentz and QED invariance. This leads to the following
dimension-six  operators
\begin{align}\label{vector:eff}
&\L \f{\alpha}{4\pi}\f{g_{ZCC}'}{4\Ld_1^2}{Z'_S}^\mu_{\,\,\nu}F^{\nu\alpha}F_{\alpha\mu}+\f{\alpha}{4\pi}\f{g_{ZCC}'}{4\wt\Ld_1^2}{Z'_S}^\mu_{\,\,\nu}\wt F^{\nu\alpha}F_{\alpha\mu}\R\cr
&+\L \f{\alpha}{4\pi}\f{g_{ZCC}'}{4\Ld_2^2}{Z'_A}^\mu_{\,\,\nu}F^{\nu\alpha}F_{\alpha\mu}+\f{\alpha}{4\pi}\f{g_{ZCC}'}{4\wt\Ld_2^2}{Z'_A}^\mu_{\,\,\nu}\wt F^{\nu\alpha}F_{\alpha\mu}\R,
\end{align}
where the  symmetric and antisymmetric 2-rank tensors are defined as $(Z_{S/A})_{\mu\nu}=\partial_\mu Z'_\nu\pm\partial_\nu Z'_\mu$.
Under   $C$ and $P$ symmetries, the vector field   transforms as
\begin{align}\label{}
C VC^{-1}=-V,\quad P V(\vec x,t)P^{-1}=(-1)^{\mu}V(-\vec x,t),
\end{align}
 and $ P\partial^\mu P^{-1}=(-1)^{\mu}\partial^\mu$. As a result
  only  $\wt \Ld_{1,2}$ term conserves the CP, so in this work we only keep them (which is consistent with the presence of  $\gamma^5$ in  Eq.~(\ref{vector:eff})).
From the effective Lagrangian, the annihilation rate is calculated to be (for illustrative purpose we only show the  $\wt\Ld_1$-related part)
   \begin{align}\label{}
(\sigma v)_{2\gamma}&\simeq\alpha^2\f{g_{Z'\chi\chi}^2g_{Z'CC}^2}{16\pi^2}\f{f_B}{64\pi }\L\f{m_\chi}{\wt\Ld_1}\R^2\f{1}{\wt\Ld_1^2}\cr
&\approx0.03\L\f{g_{Z'\chi\chi}g_{Z'CC}}{1.0}\R^2\L\f{f_B}{500}\R\L\f{m_\chi}{100\rm\,GeV}\R^2\L\f{100\rm\,GeV}{\wt\Ld_1}\R^4\rm\,pb,
\end{align}
 On the other hand, from direct calculation similar to the Ref.~\cite{Bergstrom:1997fj}, we get
   \begin{align}\label{}
(\sigma v)_{2\gamma}&\simeq N_c^2Q_c^4\L g_{Z'\chi\chi}g_{Z'CC}\R^2 \f{\alpha^2}{64\pi^3}\f{f_B}{m_\chi^2}  |{\cal A}^{1/2}(\tau)|^2.
\end{align}
This leads to the effective scale
   \begin{align}\label{}
\wt\Ld_1=m_{\wt\chi}/2Q_CN_C^{1/2}|{\cal A}^{1/2}(\tau)|^{1/2},
\end{align}
where $\tau=m_C/m_\chi$ and ${\cal A}^{1/2}(\tau)=\tau  (\arctan1/\sqrt{\tau-1})^2\gtrsim 1$ for $\tau>1$.

\section{Charged-loop porting  dark matter}

In this section we consider the scenario in the absence of resonance. Viewing from  the UV-completion level
(illustratively showed in  Fig.~\ref{2gamma}, where  a $Z_2$ symmetry can be consistently assigned on DM and charged particle),
 the operators list in the Eq.~(\ref{effective:1})
are generated in the $t-$channel. To get a large annihilation cross section without resonance enhancement,
  it is expected that there is  rather   strong coupling between DM and the charged particles. That maybe consistent with the composite dark matter. However, in the case of decaying DM scenario~\cite{Park}, a large  coupling constant can be avoided.
%%%%fig.2%%%%%%%%%%%%%%%%
 \begin{figure}[htb]
\begin{center}
\includegraphics[width=4.5in]{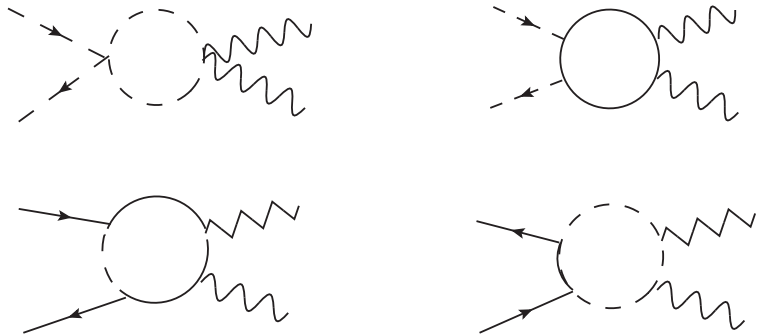}
\end{center}
\caption{\label{2gamma} Scalar/Fermionic dark matter annihilates into  $2\gamma$ via a  charged loop. $Z_2$ symmetry can be assigned to the dark matter and charged particles.}
\end{figure}
%%%%%%%%%%%%%%%%%%%%%%%%%%%%%%%%%%%%%%%%%%%%%%%%%%%%%%%%%%%%%%%%%%%

\subsection{Annihilating scenario}

In all cases, the heavy charged loop is the major  port between DM and visible sector. This is a reminiscence of the dipole dark matter theory, where the DM-photon interactions  are the leading order of DM-visible interactions.
Considering  the fermonic DM case, the effective operators up to dimension-seven should be incorporated:
\begin{align}\label{d=5}
&-\f{\ld_\chi}{2}\L \bar\chi \sigma_{\mu\nu} \chi \R  F^{\mu\nu},\quad -i\f{d_\chi}{2}\L \bar\chi \sigma_{\mu\nu}\gamma^5\chi \R  F^{\mu\nu}, \\
& \f{e}{\Ld_3^3}\bar\chi  \chi F^{\mu\nu}F_{\mu\nu},\quad i\f{e}{\Ld_4^3}\bar\chi\gamma^5 \chi F^{\mu\nu}\wt F_{\mu\nu},\label{d=7}
\end{align}
Eq.~(\ref{d=5})  are dubbed  magnetic momentum DM (MDM) and  electronic  momentum DM (EDM) respectively, with $\ld_{\chi}/d_{\chi}$
the magnetic/electric diploe momentum.
 If the DM is a Majorana fermion, Eq.~(\ref{d=5}) vanishes and we only have to consider Eq.~(\ref{d=7}). Otherwise,  we expect
\emph{all} operators listed in  Eq.~(\ref{d=5}) and Eq.~(\ref{d=7}) are comparably important,  since they are generated at  the same loop-level.  The naive dimension counting gives
\begin{align}\label{}
1/\ld_\chi\sim\f{e}{16\pi^2}\f{g_{hcc}^2}{m_C},\quad \Ld_3\sim \Ld_4\sim (m_C^2/\ld_\chi)^{1/3},
\end{align}
up to  an  overall operator coefficients  loop factors $f(m_\chi^2/m_C^2)$, whose  exact expression is very involved due to multi propagators, and we leave it for further work.

Before dealing with the  annihilating rate into gamma pair, we  consider the possible bound on operators. In spite of loop suppression, the M(E)DM has long-distance interactions, mediated by the photon, which  lead to a great  enhancement on  $\sigma_{\rm SI}$.
 Especially, the EDM has a further  $1/v^2\sim10^6$ enhancement~\cite{EDM,DelNobile:2012tx}, consequently the XENON100 tightly bounds on them. The DM-proton (only for proton by virtue of the QED mediator) cross section is
$\sigma_{\rm SI}^p=\alpha \ld_\chi^2$.  In the light of Ref.~\cite{DelNobile:2012tx}, for a 100 GeV DM, the upper bound  is roughly
\begin{align}\label{}
\ld_\chi\lesssim 10^{-19}\,\,\rm e\,cm\approx 1.5\times10^{-6}\,\, GeV^{-1},
\end{align}
which indicates   $m_C/g_{hcc}^2\gtrsim 1$ TeV. 
  In a way similar to the one given in  Section.~\ref{topW},  the  correlation between operator coefficients  leads to an estimation rate of the $\bar\chi\chi\ra\gamma\gamma$ process
  \begin{align}\label{}
(\sigma v)_{2\gamma}\sim\f{e^{2}}{64\pi }\ld_\chi^2\lesssim10^{-6}\rm\,pb,
%\approx0.08\L\f{f_B}{100}\R\L\f{100\rm\,GeV}{\Ld_1}\R^2\L\f{g_{\phi\chi\chi}'h_{\phi CC}}{5}\R^2\rm\,pb,
\end{align}
which is far below the sensitive bound. Thus the DM can not be a  Dirac particle.

Direct detection possibly  gives  a second  constraint,  no matter Dirac or Majorana DM. If the charged particles also carry color charge, then
   after replacing the  QED field-strength  with the gluon field-strength, we get the dimension-seven operators
\begin{align}\label{}
& f_G\f{\alpha_s}{4\pi} \bar\chi  \chi G_a^{\mu\nu}G^a_{\mu\nu},\quad f_{G,5}\f{\alpha_s}{4\pi} \bar\chi \gamma^5 \chi G_a^{\mu\nu}\wt G^a_{\mu\nu},
\end{align}
Again from naive dimension estimation, $f_{G,5}\sim f_G\sim 16\pi^2/e\Ld_3^3$.   Unlike the MDM or EDM operators, they  always  lead to direct detection signals. Concretely,  the first operator  gives  contribution to $\sigma_{\rm SI}^p$ in the form of  Eq.~(\ref{top:DD}), with~\cite{Jungman:1995df}
\begin{align}\label{}
\f{f_p}{m_p}\simeq -\f{2}{9}f_Gf_{TG}^p\lesssim 10^{-9}{\rm\,GeV^{-2}}/m_n,
\end{align}
which implies  the lower bound  $\Ld_3\lesssim (16\pi^2f_G/e)^{1/3}\simeq 3.7$ TeV  renders again a  very small $(\sigma v)_{2\gamma}\sim 10^{-6}$ pb. Now we can draw the conclusion: in the charged loop ported DM scenario, if we expect a large annihilation rate to gamma, XENON100 excludes both  a Dirac  DM and colored charged loops.

We would like to comment on the scalar DM case, the $\L S^\dagger\partial_\mu S\R \partial_\nu F^{\mu\nu}$ leads to contact interaction between DM and nucleon and thus safe. In fact,  such model~\cite{JMC} has been recently proposed to explain the anomaly using the scalar DM.

%We have to stress the importance of the stop-top loop induced LSP-LSP$\ra \gamma\gamma$ annihilation.
  %an effective theory irrespective of the content in the charged loop can be built utilizing the following operators (up to dimension-six):
% On our purpose,  an effective operator description is helpful to understand the DM with interaction this type.
%\begin{align}\label{}
%&i\L S^\dagger\partial_\mu S\R \partial_\nu F^{\mu\nu},\cr
%& |S|^2F^{\mu\nu}F_{\mu\nu},\quad |S|^2F^{\mu\nu}\wt F_{\mu\nu},
%\end{align}
%Similarly,

\subsection{Decaying  scenario}

The scenarios discussed previously  involve either  some tuning or strong couplings, while decaying DM gives an alternative more natural  choice.  We close the   paper by  giving  a short comment on this scenario to explain the FERMI gamma-ray line.
 Here, the  scalar DM is more or less favored. At the two-body decay level (three-body or more leads to too wide spectrum to account for the peak),  the scalar DM can decay into the gamma pair while the fermionic DM can not.
 The unique  effective operator is written as
\begin{align}\label{h}
\f{\alpha}{4\pi}\f{1}{4\Ld^2}\chi^\dagger\chi F^{\mu\nu} F_{\mu\nu},
\end{align}
but this time the  $\chi$ obtains  a TeV scale  VEV $v_\chi$ which breaks the $Z_2$ symmetry and leads to the scalar DM (the real part of $\chi$) decaying to gamma pair. To fit the data, besides a mass of DM should be around 260 GeV,   we further  need its extremely narrow branching decay width to gamma pair $\sim10^{-29}s^{-1}$~\cite{Park}. The small decay width can be achieved by lifting  the mass of charged particles running  in the loop, typically $\Ld_c\ra M_{\rm GUT}$ (see an example in~\cite{kang:decay}):
\begin{align}\label{}
\Gamma_{\chi\ra 2\gamma}=&\f{\alpha^2}{1024\pi^3}\L\f{\sqrt{2}v_\chi}{\Ld}\R^2\f{m_\chi^3}{\Ld^2}\cr
=&1.1\times10^{-29}\L\f{v_\chi}{10^3\rm\,GeV}\R^2\L\f{m_\chi}{260\rm\,GeV}\R^3\L\f{3\times10^{14}\rm\,GeV}{\Ld}\R^4 s^{-1}.
\end{align}

Some comments are in orders. Firstly, in our notation the  $\Ld$ lies  below $M_{\rm GUT}$ about two orders, but actually it may be compensated  by loop factors and large Yuawa couplings (set to unit in the above estimation). Secondly,
the relic density is a generic  problem for decaying DM, but non-thermal production such as Ref.~\cite{Kang:2010ha} using freeze-in mechanism~\cite{feeezein} may offer a solution. Last but never the least, the  130 GeV gamma-ray line in decaying DM scenario  is easily  compatible with the sharp excess in  PAMELA positron fraction~\cite{PAMELA},
which can be interpreted by a  leptonic decaying DM (saying to $e^+e^-$) with
 lifetime $\tau\sim10^{26}$s and mass around 200 GeV.
 In model building, one has to introduce relevant dimension-six operators for DM decay to leptons and adjust the  branching ratio to fit both datas.

For the fermionic DM, the Lorentz invariance force the presence of second fermion (in the final state) in the operator of decay.  As a case in point, in the (lepton number violating) $R-$parity violating SUSY, the gravitino has the following  two body  decay modes~\cite{Ibarra:2007wg}:
 \begin{align}\label{}
 \wt G\ra \nu+\gamma, \quad W^++\ell^-,\quad Z^0+\nu,
\end{align}
where  gravitino mass is $m_{\wt G}=250$ GeV and the branching ratios  are respectively given by 0.03, 0.69 and 0.28.
Although it fails to explain both PAMELA and FERMI gamma-ray line, the first mode can explain the latter given proper decay width. Of particular interesting,
 the neutrino (from the third mode) and gamma signal appear simultaneously, and the accompanied  neutrino signal may be detected and thus provides  a complementary detect method for this scenario.

\section{Conclusion and discussion}\label{conclusion}

The   gamma-ray line from DM  annihilating in the galaxy center  generically is  well below the present detectable  level.
However,  once the observation, from it we are able to extract very important information of the DM properties/dynamics.
  In this work, we present a minimal  effective theory framework to  understand the anomalously bright gamma-ray line from dark matter activity:
  \begin{itemize}
    \item  In the EFT  large annihilation rate is ascribed to operator coefficients with
resonant or strong coupling enhancement.
    \item Due to the XENON100 bound, Dirac DM or  colorful charged particles are ruled out in models with only strong couplings.
    \item If the charged particle in the loop carry $SU(3)_C$ charge, the di-gluon annihilation  mode is about one order larger than the di-photon mode, that may properly  account for the  relic density.
        \item The SM-like  Higgs may share the same charged loop, and therefore provide a source of Higgs  di-photon excess at the LHC.
  \end{itemize}
%   We that can interpret the tentative  130 GeV gamma-ray line. Top-window model is also proposed to explain it.
  %Bases on quasi-effective analysis, we investigate the origin of large rate consistent with other constraints:
Applying  the general analysis to the NMSSM, that  is proved to accommodate neutralino LSP with large annihilation rate into di-photon
and interpret the tentative  130 GeV gamma-ray line. Top-window model is also proposed to explain it.

Although not the central points of this work, we would like to end up by commenting its very promising collider  detection prospect, if the 130 GeV gamma-ray line from DM activity will be confirmed. In light our general analysis in the text,  at the LHC or Tavertron  (but beyond LEP) one can expect new light color-singlet charged particle $C$
can be produced: $q\bar q\ra C^\dagger C$. While beyond the 130 GeV line and consider more wide scope, the LHC could put very  strong exclusion on the model with  colored loop.

\section*{Acknowledgement}

We  thank  Wanlei Guo, Da Huang, Ling-Fong Li, and Jing Shu for helpful discussion. And we greatly appreciate Zheng Sun for reading the draft.
This research was supported in part by the Natural
Science Foundation of China under grant numbers
10821504, 11075194, and 11135003, and by the United
States Department of Energy Grant Numbers DE-FG03-
95-Er-40917.

\section*{Note added}

In the completion of this work, we note the appearance of  work~\cite{Chu:2012qy}, they also note the importance of the di-gluon mode in the determining of
DM relic density and more relevant phenomenologies are discussed there. The Ref.~\cite{Das:2012ys}  specifically studies the 130 GeV gamma-ray line from LSP annihilation in the NMSSM, taking a quite similar   idea to ours. We greatly thank  the authors for sending us the updated version of  NMSSMtools in which the CP-odd mass correctly adopts the running mass. Using it we re-scan our region, and find  different results  than ~\cite{Das:2012ys}.

\appendix

\section{Effective vertex from the charged loop}\label{phigamma2}

\begin{figure}[hbtp]
\begin{center}
\includegraphics[scale=1.0]{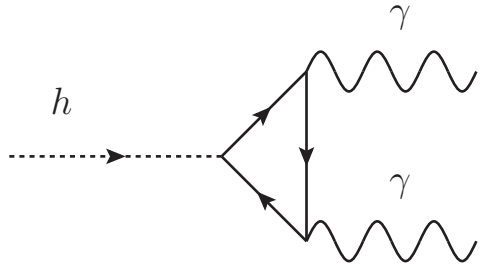}
\caption{fermion loop induced neutral scalar decay}\label{Fig:abc}
\end{center}
\end{figure}

\subsection{$\phi\ra2\gamma$}
In this Appendix we present the procedure of  calculating  the coefficients of the   effective vertex used in the Section~\ref{res:scalar}, quoted for convenience:
 \begin{align}\label{phi2gamma1}
\alpha\f{ h_{\phi CC}}{4\pi}\f{1}{4\Ld_1} \phi_h F_{\mu\nu}F^{\mu\nu},\quad
\alpha\f{ h_{\phi CC}}{4\pi}\f{1}{8\Ld_2} \phi_A F_{\mu\nu}\wt F^{\mu\nu}.
\end{align}
The calculation is similar to the case of the  Higgs with general couplings. For definiteness, here we focus on the process depicted in the Fig.~\ref{Fig:abc}, where CP-even resonance $\phi_h$ decays to gamma pair with single  charged fermion running in the loop.
  At the one hand, the direct calculation of  decay width gives %The partial decay widths of scalar $h$ into two photon is given by
\begin{eqnarray}\label{h2ga}
\Gamma\left(h\rightarrow\gamma\gamma\right) = \frac{\alpha^2m_{h}}{256\pi^3} \left|
2N_cQ_C^2 h_{\phi CC}\mathcal{A}^{h}_{1/2}\left(\tau\right)\right|
^2,\label{CPeven}
\end{eqnarray}
with $N_c=3$ the color factor and $Q_C$ the electronic  charge of $C$. The loop function  $\mathcal{A}(\tau)$ is
\begin{eqnarray}
\mathcal{A}_{1/2}^h(\tau) =& \frac{2}{\tau^{{3}/{2}}}\left[\tau+(\tau-1)f(\tau)\right],\\
f(\tau) =& \left\{ \begin{array}{cc}
                \arcsin^2\sqrt{\tau}                    & \tau \leq 1 \\
                -\frac{1}{4}\left[\log\frac{1+\sqrt{1-\tau^{-1}}}{1-\sqrt{1-\tau^{-1}}}\right]^2  & \tau > 1
                 \end{array} \right.
\end{eqnarray}
only depending of the ratio $\tau = m_{\phi_h}^2/4m_{C}^2$.
In  general, the  $\phi_h$ can be off shell and therefore the replacement  $m_{\phi_h}^2\ra P^2$, with $P$ the four-momentum of $\phi_h$.
For the CP-odd resonance $\phi_A$, the decay width takes the same form as the Eq.~(\ref{h2ga}) but the loop function is different:
 \begin{eqnarray}
\mathcal{A}^{A}_{1/2} = 2\tau^{-1/2}f(\tau).
\end{eqnarray}
On the other hand, from the effective operators in  Eq.~(\ref{phi2gamma1}) one
 can  calculate
\begin{eqnarray}
\Gamma\left(\phi_h\rightarrow\gamma\gamma\right) =\frac{\alpha^2h_{\phi CC}^2}{1024\pi^3\Lambda_{1}^2}m_{\phi_h}^3, \quad
\Gamma\left(\phi_A\rightarrow\gamma\gamma\right)= \frac{\alpha^2h_{\phi CC}^2}
{1024\pi^3\Lambda_{2}^2}m_{\phi_A}^3.\label{Adecay}
\end{eqnarray}
So, comparing Eq.~(\ref{CPeven}) and  Eq.~(\ref{Adecay}), we eventually  get the effective suppressing scales
\begin{eqnarray}\label{QEDLd}
{\Lambda_{1}} = {m_{\chi}}/|2N_c Q_C^2\mathcal{A}^{h}_{1/2}|,\quad {\Lambda_{2}} = {m_{\chi}}/|2N_c Q_C^2\mathcal{A}^{A}_{1/2}|,
\end{eqnarray}
where $m_\phi=2m_\chi$ has been used. Those expressions   justify the statement  that  the $\Ld_{1,2}$ can be much below the weak scale (or $m_\chi$).

\begin{figure}
\includegraphics[scale=0.9]{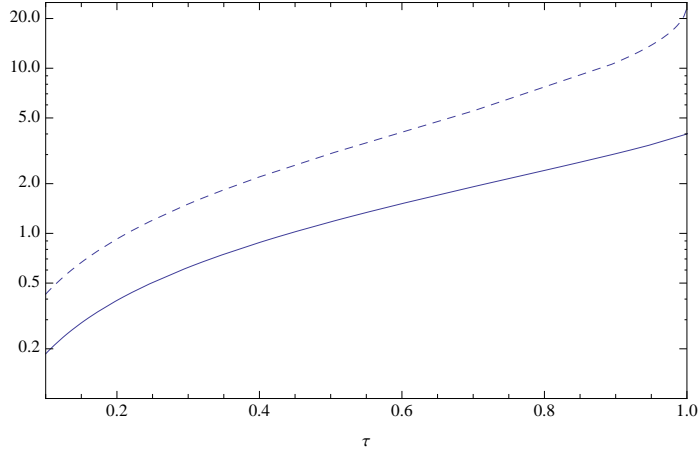}
\caption{The loop function $|{\cal A}|^2$ varies as variable $\tau=m_\phi^2/4m_C^2$. Dashed line: for the CP-odd $\phi$;  Solid line: for CP-even $\phi$. The former is always several times of  the latter.}\label{CPoddfactor}
\end{figure}

\subsection{$\phi\ra 2G$,\,\quad $\phi\ra 2Z$}

There are two other  effective vertex require  attention. First,
if the charged particle also carries color charge, then the resonance  can decay into two gluons. Repeating  the procedure dealing with the vertex $\phi_h FF$ and $\phi_A F\wt F$, the corresponding   coefficients  parameterized exactly  as the Eq.~(\ref{phi2gamma1}) except for the replacement  $\alpha\ra\alpha_s$ and $\Ld_{i}\ra\Ld_{i,s}$, we are able to get the precise effective operators.
  It is straightforward  to get the gluonic partial
decay width from direct and effective calculations:
\begin{eqnarray}
\Gamma\left(\phi_h\rightarrow gg\right)& = \frac{\alpha_{s}^2m_{\phi_h}}{32\pi^3}|
h_{\phi CC}\mathcal{A}_{1/2}^{h}|^2=\frac{\alpha_{s}^2h_{\phi CC}^2}
{128\pi^3\Lambda_{1}^3}m_{h}^3,\\
\Gamma\left(\phi_A\rightarrow gg\right)& = \frac{\alpha_{s}^2m_{\phi_A}}{32\pi^3}|
h_{\phi CC}\mathcal{A}_{1/2}^{A}|^2=\frac{\alpha_{s}^2h_{\phi CC}^2}
{128\pi^3\Lambda_{2}^3}m_{A}^3.\label{CPoddgluon}
\end{eqnarray}
Then  we obtain the effective scales
\begin{eqnarray}\label{Lds}
{\Lambda_{1,s}}= {m_{\chi}}|/|\mathcal{A}_{1/2}^{h}|,\quad {\Lambda_{2,s}}= {m_{\chi}}|/|\mathcal{A}_{1/2}^{A}|.
%\sum_{f}Y_{f}\mathcal{A}_{1/2}^{h}|^2,\quad
%\frac{h_{\phi CC}^2}{\Lambda_{2}^2} = \frac{4}{m_{A}^2} |
%\sum_{f}Y_{f}\mathcal{A}_{1/2}^{A}|^2
\end{eqnarray}
a few times of the values given in the Eq.~(\ref{QEDLd}).

Now we turn  attention to the di-$Z$ mode. Compared to the di-phonon mode, the  difference lies in the additional terms in the
 polarization vector substitution:  $\epsilon_\mu\epsilon^*_\nu\ra g_{\mu\nu}-p_1^\mu p_2^\nu/m_Z^2$. Presumably the resulting change is suppressed by the small parameter  $m_Z^2/m_C^2$ (confirmed by the $h \rightarrow Z\gamma$ result~\cite{Djouadi:2005gi}), then we are justified to ignore this effect at the leading order and approximately have
%Specially when we mainly consider the contribution of heavy fermions, the difference is neglectable even though
%the mass of $Z$ boson. So we consider the loop induced decay of $\phi \rightarrow ZZ$ and focus on the
%case of heavy fermion entering into loop. Then we can approximatively calculate the partial decay width
%of $h \rightarrow ZZ$ by virtue of comparing coupling of $\overline{\psi}\gamma^{\mu}\psi A_{\mu}$ and
\begin{eqnarray}
\frac{\Gamma(h \rightarrow ZZ)}{\Gamma(h \rightarrow \gamma\gamma)} \sim
\frac{g^4\hat{v}_{f}^4}{e^4Q^4_C}\L1 - 4\frac{M_{Z}^2}{m_{\phi}^2}\R^{3/2},
\end{eqnarray}
the phase space suppressing factor is about 0.3 for $m_\phi=260$ GeV, largely it cancels the enhancement from coupling ratio.
Generally the vector coupling is of the form
$\hat v_f\overline{\psi}\gamma^{\mu}\psi
Z_{\mu}$ with  $\hat{v}_{f} = T^{3}_C - 2Q_C\sin^2\theta_{W}$, here $T_C^3$  is the isospin quantum number  of $C$.

\end{document}